\documentclass[reprint,amsmath,amssymb,aps,pre,superscriptaddress
]{revtex4-1}
\usepackage{blindtext}
\usepackage{comment}
\usepackage{tikz}
\usepackage{subfigure}
\usepackage{graphicx}
\usepackage{dcolumn}
\usepackage{mathtools,amsmath}
\usepackage{amssymb}
\usepackage{float}
\usepackage{graphicx}
\usepackage{subfigure,bm}
\usepackage[export]{adjustbox}
\usepackage{type1cm}
\usepackage{eso-pic}
\usepackage{color}
\usepackage{dcolumn}
\newcolumntype{x}[1]{>{\centering\arraybackslash\hspace{0pt}}p{#1}}

\begin{document}

\preprint{APS/123-QED}
\title{Nonmonotonic dependence of polymer glass mechanical response\\on chain bending stiffness}
\author{Christopher Ness}
\author{Vladimir V. Palyulin}
\author{Rico Milkus}
\affiliation{Department of Chemical Engineering and Biotechnology, University of Cambridge, Cambridge CB3 0AS, United Kingdom}

\author{Robert Elder}
\affiliation{U.S. Army Research Laboratory, Aberdeen Proving Ground, MD, United States}
\affiliation{Bennett Aerospace, Inc., Cary, NC, United States}
\author{Timothy Sirk}
\affiliation{U.S. Army Research Laboratory, Aberdeen Proving Ground, MD, United States}

\author{Alessio Zaccone}
\affiliation{Department of Chemical Engineering and Biotechnology, University of Cambridge, Cambridge CB3 0AS, United Kingdom}
\pacs{Valid PACS appear here}
\date{\today}

\begin{abstract}
We investigate the mechanical properties of amorphous polymers by means of coarse-grained simulations and nonaffine lattice dynamics theory. A small increase of polymer chain bending stiffness leads first to softening of the material, while hardening happens only upon further strengthening of the backbones.  This nonmonotonic variation of the storage modulus $G'$ with bending stiffness is caused by a competition between additional resistance to deformation offered by stiffer backbones and decreased density of the material due to a necessary decrease in monomer-monomer coordination. This counter-intuitive finding suggests that the strength of polymer glasses may in some circumstances be enhanced by softening the bending of constituent chains.
\end{abstract}

\maketitle

\paragraph*{Introduction}
The study of polymer dynamics has been at the heart of soft matter research for decades, yet a comprehensive theoretical basis that links monomer chemistry to mechanical properties remains under development~\cite{doi1988theory,Rubinstein2003}.
Polymers below their glass transition temperature, which find application in everyday consumer goods and high-technology material applications, pose a particular challenge as understanding their properties further requires an assimilation of glassy dynamics, itself a topic of ongoing debate~\cite{hoy2011understanding}. 

Throughout the historical development of polymer physics, it has proven constructive to consider two idealised linear polymer models: freely-jointed, in which chains are assumed to comprise random walks of fixed step length with no monomer interactions; and freely-rotating, in which the angle formed by three consecutive monomers is strictly fixed but the monomers are otherwise unconstrained. Here we explore the mechanical properties of polymer glasses between these limits as the monomer motions become increasingly constrained by a bending penalty. We further enforce excluded volumes around individual monomers.

It is already established that increasing the number of constraints on particles in a many-body system reduces the critical coordination, and hence the critical density, at which the system achieves marginal stability~\cite{he1985elastic}. This has been apparent in granular systems for some time, when comparing frictionless to frictional packings~\cite{song2008phase}. Indeed, constraint-counting arguments underpin recent theories for shear thickening in athermal suspensions~\cite{wyart2014discontinuous}. An analogy between friction in granular systems and bending in polymers has been proposed theoretically~\cite{rosch2013exploring} and in experiments on `granular polymers'~\cite{zou2009packing} and is a promising lead towards unifying the understanding of marginal stability across a surprisingly broad class of soft matter systems~\cite{karayiannis2009contact,lopatina2011jamming}.

The introduction of bending constraints in bead-spring polymer chains is expected to reduce the critical coordination $Z_g$, $i.e.$ the sum of inter- and intra-chain interactions at the glass transition, and, therefore, the critical density~\cite{hoy2017jamming}. Such a density reduction is reminiscent of the role of plasticising additives~\cite{papakonstantopoulos2016controlling}, designed to \emph{reduce} the mechanical strength of the material by increasing the free volume. By contrast, one might expect enhanced bending stiffness to increase the strength of the bulk material. The question remains, therefore, what overall effect the introduction of such constraints has on the mechanical properties of glassy polymers.

In this Rapid Communication we show using simulations and theory that the competing effects of increasing backbone strength and increasing free volume lead to nonmonotonic behaviour of the shear modulus of glassy polymers as a function of bending stiffness. This finding offers a connection between monomer chemistry and polymer glass rheology, demonstrating that manipulating bending constraints at the monomer level can have nontrivial influence on the bulk mechanical properties of the material.

\begin{figure*}
{\includegraphics[trim = 0mm 5mm 0mm 0mm, clip,width=0.65\textwidth]{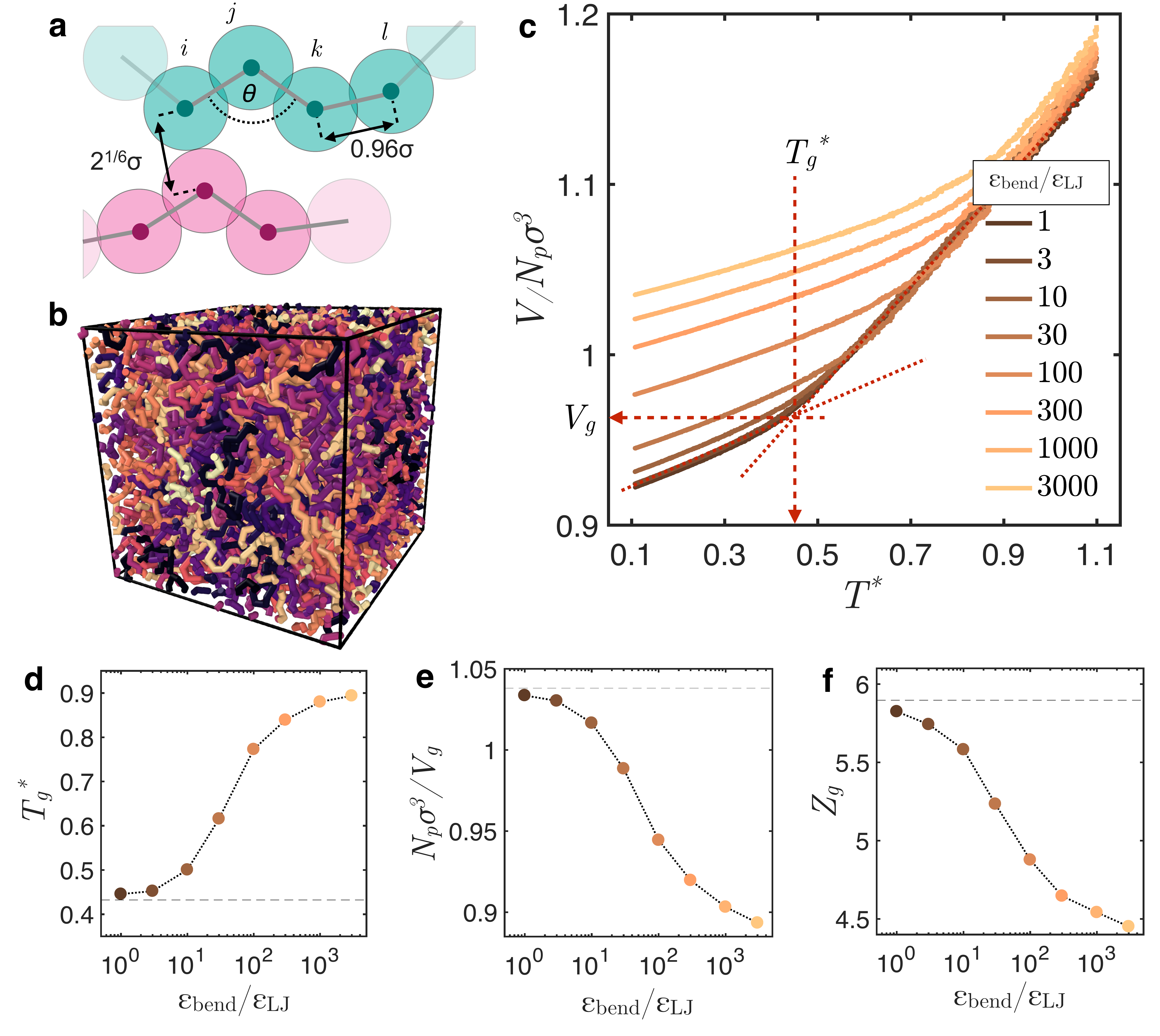}}
\caption{
Entry into the glassy state and its structural properties for $\theta_0 = 109.5^\circ$.
(a) Sketch of polymer chain illustrating the angle $\theta$ and rest positions for LJ ($2^{1/6}\sigma$) and FENE ($0.96\sigma$) interactions.
(b) Snapshot of glassy polymer in periodic box~\cite{stukowski2009visualization}.
(c) The decrease of volume associated with decreasing $T^*$ at fixed pressure, for several values of $\varepsilon_\mathrm{bend}/\varepsilon_\mathrm{LJ}$. We approximate the low- and high-temperature dependences as linear, and take their intersection to occur at $T^* = T_g^*$.
(d) Variation of glass transition temperature $T^*_g$ with $\varepsilon_\mathrm{bend}/\varepsilon_\mathrm{LJ}$.
(e) Variation of $N_p\sigma^3/V_g$, the density at $T^*_g$, with $\varepsilon_\mathrm{bend}/\varepsilon_\mathrm{LJ}$.
(f) Variation of $Z_g$, the coordination number at $T^*_g$, with $\varepsilon_\mathrm{bend}/\varepsilon_\mathrm{LJ}$.
Dashed line in (d)-(f) indicates $\varepsilon_\mathrm{bend}=0$.
}
\label{figure1}
\end{figure*}

\begin{figure*}
{\includegraphics[trim = 0mm 0mm 46mm 0mm, clip,width=0.95\textwidth]{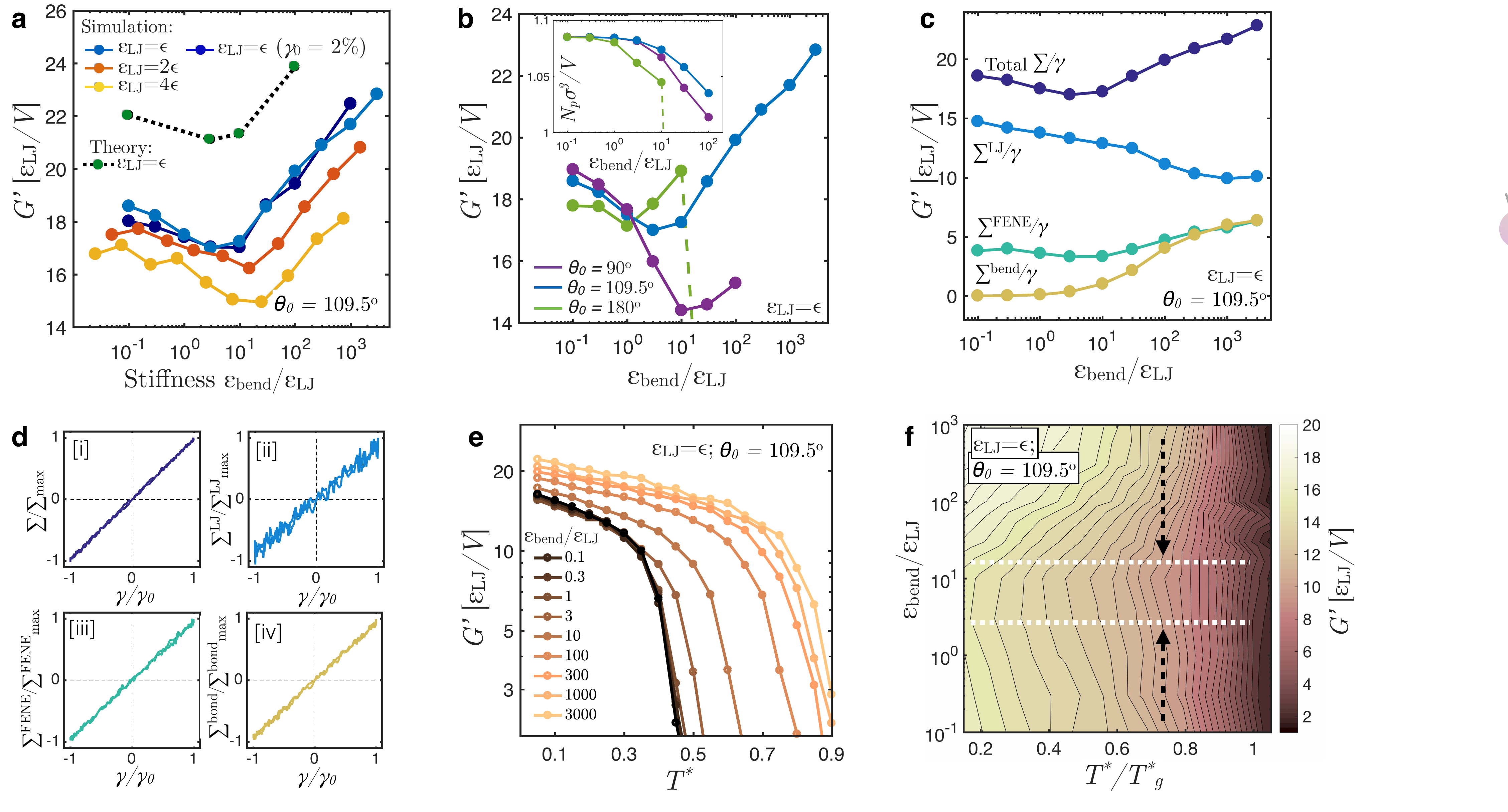}}
\caption{
Nonmonotonic mechanical response of a polymer glass as a function of chain bending stiffness.
(a) Elastic modulus $G'$ as a function of $\varepsilon_\mathrm{bend}/\varepsilon_\mathrm{LJ}$ for three values of $\varepsilon_\mathrm{LJ}$. We used five realisations and found the variation between realisations to be smaller than the marker size.
(b) Elastic modulus $G'$ as a function of $\varepsilon_\mathrm{bend}/\varepsilon_\mathrm{LJ}$ for three values of $\theta_0$. Inset: decreasing density with increasing $\varepsilon_\mathrm{bend}/\varepsilon_\mathrm{LJ}$.
(c) Elastic modulus contributions from LJ, FENE and angular potentials for $\varepsilon_\mathrm{LJ} = \epsilon$ and $\theta_0=109.5^\circ$.
(d) Lissajous-Bowditch plots showing linear elastic stress contributions [i] the total $\Sigma$; [ii] $\Sigma^\mathrm{LJ}$; [iii] $\Sigma^\mathrm{FENE}$; and  [iv] $\Sigma^\mathrm{bend}$, each rescaled by their maximal values. Strains are rescaled by the amplitude $\gamma_0$.
(e) Temperature dependence of $G'$ across a range of bending stiffnesses $\varepsilon_\mathrm{bend}/\varepsilon_\mathrm{LJ}$.
(f) Contour plot showing $G'$ as a function of rescaled temperature $T^*/T^*_g$ and $\varepsilon_\mathrm{bend}/\varepsilon_\mathrm{LJ}$. Dashed black arrows indicate decreasing $G'$; dotted white lines show region of minimal $G'$.
}
\label{figure2}
\end{figure*}

\paragraph*{Simulation details}
A non-overlapping random-walk algorithm is used to generate initial loose configurations of $N_p=10^4$ monomers, in chains of length $10^2$.
For each monomer in our system we use \texttt{LAMMPS}~\cite{plimpton1995fast} to solve the Langevin equation
with coefficient of friction $1/\xi$ and random forces ${f}_B(t)$ satisfying $\langle f_B(t)f_B(t')\rangle = 2mk_BT\delta(t-t')/\xi$ at time $t$.
Monomers of uniform mass $m$ interact through potentials $U$ given by the Kremer-Grest model~\cite{kremer1990dynamics}, comprising a Lennard-Jones potential $U^\mathrm{LJ}(r) = 4\varepsilon_\mathrm{LJ} \left[\left(\frac{\sigma}{r}\right)^{12} - \left(\frac{\sigma}{r}\right)^6 - \left(\left(\frac{\sigma}{r_c}\right)^{12} - \left(\frac{\sigma}{r_c}\right)^6\right) \right]$ with depth $\varepsilon_\mathrm{LJ}$ and rest length $2^{1/6}\sigma$ acting between monomer pairs within a cut-off range $r_c=2.5\sigma$ and a finitely extensible nonlinear elastic potential $U^\mathrm{FENE}(r) = -0.5\varepsilon_\mathrm{FENE} R_0^2 \ln \left[1 - \left(\frac{r}{R_0}\right)^2\right]$ with maximal length $R_0$ and emerging rest length $\approx0.96\sigma$ acting between sequential monomer pairs along each chain~\footnote{The bidispersity introduced by differing rest lengths of LJ and FENE is sufficient to suppress crystallisation throughout~\cite{zaccarelli2009crystallization,nguyen2015effect}}. $\varepsilon_\mathrm{LJ}$ sets the LJ energy scale and $\varepsilon_\text{FENE}$ is the bond energy scale where $\varepsilon_\mathrm{FENE}/\varepsilon_\mathrm{LJ} = 30$. With reference to fundamental units of mass $\nu$, length $d$, and energy $\epsilon$, we set $\sigma=1$ and $m=1$, giving a time unit of $\tau=\sqrt{m\sigma^2/\varepsilon_\mathrm{LJ}}$, and we set $\xi=100\tau$. We define a third energy associated with chain bending given by
$U^{\mathrm{bend}}(\theta) = \varepsilon_\mathrm{bend}[1-\cos(\theta-\theta_0)]$
for energy scale $\varepsilon_\mathrm{bend}$ and rest angle $\theta_0$. The angle $\theta$ is formed between consecutive monomer triplets along each linear chain, Fig~\ref{figure1}a.
A dissipative timescale emerges as $m\sigma^2/\xi\varepsilon_\mathrm{LJ}$, and a thermal timescale emerges as $m\sigma^2/\xi k_BT$. The state of our system, i.e. whether it is in the melt or glassy state, is given simply by the ratio of these timescales, as $T^* = k_BT/\varepsilon_\mathrm{LJ}$~\footnote{A comprehensive description of our simulation model is provided in the Supplementary Material}. A snapshot of the polymer glass is given in~Fig.~\ref{figure1}b.

\paragraph*{Decreasing density with $\varepsilon_\mathrm{bend}/\varepsilon_\mathrm{LJ}$}
Using periodic boundaries we equilibrate the system in a melted state at $T^*=1.2$, maintaining zero external pressure using a Nose-Hoover barostat. We then cool the system by decreasing $T^*$ at rate $1/\tau_c$, with $\tau_c \sim \mathcal{O}(10^5)\tau$. Results are presented in Fig.~\ref{figure1} for $\theta_0=109.5^\circ$. The system undergoes a decrease in volume $V$ as it is cooled, with a change of slope at $T^*=T_g^*$,~Fig.~\ref{figure1}c~\cite{han1994glass}. 
The coordination of the system is quantified by counting all neighbouring monomers that are within the repulsive part of the Lennard-Jones, i.e.:
$
Z = \frac{N_p\sigma^3}{V}\int_0^{2^{1/6}\sigma} g(r)4\pi r^2dr \text{}
$
where $g(r)$ is the monomer-monomer radial distribution function.
The glass transition occurs at $T^*=T_g^*$, where $V=V_g$ and $Z=Z_g$.
As expected~\cite{strobl1997physics}, $T_g^*$ increases with $\varepsilon_\mathrm{bend}$, with apparent limiting values occurring for $\varepsilon_\mathrm{bend}/\varepsilon_\mathrm{LJ}\to0$ and $\varepsilon_\mathrm{bend}/\varepsilon_\mathrm{LJ} > 10^3$ (Fig.~\ref{figure1}d), while the associated density $N_p\sigma^3/V_g$ (Fig.~\ref{figure1}e) and coordination $Z_g$ (Fig.~\ref{figure1}f) decrease. $Z_g$ varies from 5.9 to 4.4, close to the expected values when transitioning from a purely central force network to one bound by bending constraints~\cite{he1985elastic,zaccone2013disorder}. A value closer to $Z_g=4$ is expected for chain lengths $\gg10^2$ and for $\varepsilon_\mathrm{bend}/\varepsilon_\mathrm{LJ} \to \infty$, while further constraints such as torsional rigidity are expected to lead to further reduction~\cite{he1985elastic}. Thus, adding constraints to the monomers reduces the critical coordination and density of the system. The trends in Figs.~\ref{figure1}e,f remain the same at any fixed $T^*<T^*_g$; the shearing simulations described below are run at $T^*=10^{-3}$ for comparison with athermal theory and at higher temperatures to test the robustness of the nonmonotonic response near $T^*_g$. Further structural description is given in Fig S1.

\paragraph*{Nonmonotonic dependence of $G'$ on $\varepsilon_\mathrm{bend}/\varepsilon_\mathrm{LJ}$}
The storage modulus $G'$ is obtained for bending stiffnesses in the range $\varepsilon_\mathrm{bend}/\varepsilon_\mathrm{LJ} = 0.01 \to 3000$ and rest angles $\theta_0=90^\circ$, $109.5^\circ$ and $180^\circ$ by two means.  
In the first, we use dynamic simulation to apply an oscillatory shear deformation to the system at $T^* = 10^{-3}$ and zero external pressure, with strain amplitude $\gamma_0 = 1\%$ and period $200\tau$. For these parameters the system remains in the linear elastic regime.
From the potentials $U^\mathrm{LJ}$, $U^\mathrm{FENE}$ and $U^\mathrm{bend}$ described above, we obtain per-monomer forces as, \emph{e.g.}, $f^\mathrm{LJ} = -dU^\mathrm{LJ}/dr$ and compute shear stresses in $xy$ (with velocity $x$, gradient $y$, vorticity $z$) according to ${\Sigma}^\mathrm{LJ} = \frac{1}{V}\sum_{n = 1}^{N^\mathrm{LJ}} {r}_{{x,n}} f^{\mathrm{LJ}}_{y,n}\text{,}$ where $N^\mathrm{LJ}$ represents the total number of LJ interactions and $r$ is the vector between interacting monomers, and similarly for FENE~($\Sigma^\mathrm{FENE}$) and bending~($\Sigma^\mathrm{bend}$) interactions. We take the total $\Sigma$ and compute $G'$ from the linear oscillatory stress response in the usual way after $\mathcal{O}(10^2)$ cycles.

In the second, we use the nonaffine lattice dynamics formalism~\cite{lemaitre2006sum,zaccone2011approximate,damart2017theory} to theoretically predict the zero-temperature elastic response from the amorphous structure. The modulus comprises an affine term $G_A$~\cite{born1954dynamical} and a nonaffine correction that originates in the lack of local inversion symmetry of the polymer glass. From the interaction potentials and particle coordinates, we obtain the affine contribution to the elastic modulus as
$G_A = \frac{1}{V}\frac{\partial^2U}{\partial\gamma^2} \bigg\rvert_{\gamma\to0} \text{,}$
where $U$ is the overall interaction potential energy and $\gamma$ is the strain amplitude. To obtain the nonaffine contribution, we first construct the Hessian matrix $H_{ij}$ for the system at a given configuration as the second derivative of the energy following Ref~\cite{lemaitre2006sum}, where the entries can in general be written as
\begin{equation}
\frac{\partial^2U(z)}{\partial r^a_n\partial r^b_m} = \underbrace{\frac{d^2U(z)}{dz^2}}_{\text{stiffness}}\frac{\partial z}{\partial r^a_n}\frac{\partial z}{\partial r^b_m} + \underbrace{\frac{dU(z)}{dz}}_{\text{tension}}\frac{\partial^2 z}{\partial r^a_n \partial r^b_m} \text{.}
\end{equation}
Here $z$ represents a generic argument that in practice is represented by either the monomer-monomer separation $r$ or the angle $\theta$; we give a detailed form of the corresponding matrix entries in the SI. $H_{ij}$ thus includes stiffness and tension contributions from Lennard-Jones, FENE and angular potentials~\cite{van2006expressions}. The eigenvalue problem
$\omega_k^2m{\bm e}_i^k = \sum_jH_{ij}{\bm e}_j^k$
is then solved directly, after which we compute the storage modulus as
\begin{equation}
G'(\Omega) = G_A - \mathrm{Re}\left(\frac{1}{V}\sum_k \frac{\Gamma(\omega_k)}{m\omega_k^2-m\Omega^2 - i\Omega\nu}\right) \text{,}
\end{equation}
where $\Gamma(\omega_k)$ is the affine force field correlator, ${\bm e}_i^k$, ${\bm e}_j^k$ are eigenvectors and the sum is over the $k$ eigenvalues of the system.

In Fig.~\ref{figure2}a we present $G'$ as a function of bending stiffness from both simulation and theory, for $\theta_0=109.5^\circ$. Shown are results for three values of $\varepsilon_\mathrm{LJ}$. We verified that our results are valid throughout the linear elastic regime by repeating the $\varepsilon_\mathrm{LJ}=\epsilon$ calculations at $\gamma_0=2\%$. There is clear nonmonotonic dependence of $G'$ on $\varepsilon_\mathrm{bend}/\varepsilon_\mathrm{LJ}$ in all cases, with a minimum in $G'$ occurring at $2<\varepsilon_\mathrm{bend}/\varepsilon_{LJ}<20$. The theoretical prediction provides a strong qualitative match to the simulation result at $\varepsilon_\mathrm{LJ}=\epsilon$, also showing nonomontonic behaviour. In the present article, we limit our discussion of the theoretical approach to its corroboration of the simulation result. Future works will focus on the detailed interpretation of the arising features of the density of vibrational states. When expressed in units of $\epsilon/V$ (Fig S2), there is a strong increase of $G'$ with $\varepsilon_\mathrm{LJ}$ as expected. In units of $\varepsilon_\mathrm{LJ}/V$, though, $G'$ collapses with a small offset for all $\varepsilon_\mathrm{LJ}$, as expected due to the decreasing \emph{relative} contribution from FENE bonds in each case.
In Fig.~\ref{figure2}b we present $G'$ at three rest angles $\theta_0$. Nonmonotonic behaviour is recovered in each case. For $\theta_0=90^\circ$, we observe an enhanced minimum, with a substantial reduction in $G'$ of approximately 25\%, correlated with a consequent decrease in density relative to $\theta_0=109.5^\circ$, Fig.~\ref{figure2}b Inset. A shallow minimum is also observed for $\theta_0=180^\circ$, though at larger values of $\varepsilon_\mathrm{bend}/\varepsilon_\mathrm{LJ}$ individual chains become rod-like, at which point both the density and $G'$ of the material have anomalous behaviour (Fig S4).

To eludicate the origin of the minimum in $G'$, we decompose the contributions $\Sigma^\mathrm{LJ}$, $\Sigma^\mathrm{FENE}$, $\Sigma^\mathrm{bend}$ from the simulation result for $\theta_0=109.5^\circ$ and $\varepsilon_\mathrm{LJ}=\epsilon$, Fig.~\ref{figure2}c, verifying that each remains linearly elastic, Fig.~\ref{figure2}d. Consistent with the decrease of $N_p\sigma^3/V_g$ and $Z_g$ with increasing stiffness (Fig~\ref{figure1}e,f), we find a steady decrease in $\Sigma^\mathrm{LJ}$ as stiffness is increased. The increasingly rigid built-in three body correlations arising from increasing $\varepsilon_\mathrm{bend}/\varepsilon_\mathrm{LJ}$ necessitate a smaller number of pairwise monomer interactions for marginal stability, which can be achieved at a lower density, or equivalently at a higher free volume. As such the stress contribution from Lennard-Jones interactions (which is proportional to the packing density) decreases monotonically. Since $\Sigma^\mathrm{LJ}$ is the dominant contribution, this corresponds initially to an overall drop in $G'$. As expected, though, we find a monotonic increase in $\Sigma^\mathrm{bend}$ as stiffness is increased, as deformation requires an increasing energy input to move three-body configurations away from their resting positions. There is a minor nonmonotonicity observed in the FENE contribution to $G'$, with the minimum being attributable to the removal of LJ interactions allowing minor relaxations of FENE bonds to their resting positions. This magnitude of this effect is, though, largely outweighed by the other contributions.

We next test the robustness of the nonmonotonic behaviour away from the low temperature limit, as the glass transition temperature $T^*_g$ is approached from below. A plot of $G'$ as a function of temperature is given in~Fig.~\ref{figure2}e. For low temperatures, the shear modulus decreases slowly with increasing temperature, until a critical value is reached at which point the mechanical rigidity is lost~\cite{zaccone2013disorder}. Rescaling the temperature axis with the appropriate values of $T^*_g$ (obtained from Fig~\ref{figure1}c), Fig~\ref{figure2}f, we find a good collapse of the loss of rigidity $G'$ as $T^*\to T^*_g$. Similarly, we find the minimal $G'$ occurring in the same range of $\varepsilon_\mathrm{bend}/\varepsilon_\mathrm{LJ}$ as in Fig~\ref{figure2}a across temperatures, highlighted by white dotted lines in Fig~\ref{figure2}f. The nonmonotonic behaviour of $G'$ thus remains even very close to the glass transition. This raises the question of the mechanism by which marginal stability is achieved in semiflexible polymers at $T_g^*$, which might extend recent work in the $T=0$ limit by Ref~\cite{hoy2017jamming}. The values of $T_g^*$ and $Z_g$ vary monotonically with chain bending stiffness between asymptotic limits (Fig.~\ref{figure1}d,f), yet the mechanical strength at the glass transition retains a minimum for intermediate $\varepsilon_\mathrm{bend}/\varepsilon_\mathrm{LJ}$.

In general, therefore, one might expect that any chemical change that decreases the monomer-monomer coordination of the system, i.e. adding bending constraints or frustrating packing by inclusion of plasticisers, will result in a decrease in the contribution to $G'$ from non-bonding interactions (represented here as Lennard-Jones). We have demonstrated here with the $\theta_0=90^\circ$ case that this might be `designed for' in practice by adjusting the rest angles of linear chains to enhance this decrease. Conversely, it follows trivially that increasing bending stiffness of polymer chains will generally increase the $G'$ contribution from angular potentials.

The two contributions to the storage modulus $G'$ from non-bonding and bending interactions thus have opposite responses to increases in chain bending stiffness.
As a result, there is a competition between these contributions that leads to an overall nonmonotonic dependence of $G'$ on bending stiffness, with there being a minimum in $G'$ at  $2<\varepsilon_\mathrm{bend}/\varepsilon_{LJ}<20$.
Parameter exploration in $\varepsilon_\mathrm{LJ}$ and $\theta_0$ demonstrate that both the depth and location of the minimum in $G'$ can be tuned by manipulating monomer chemistry, suggesting ways in which one might exploit or suppress the nonmonotonicity.
Together, these findings predict that nonmonotonicity in $G'$ is a generic feature across glassy polymeric materials.

Given the monomer chemistry of some novel polymeric system, one might use ab-initio computations to derive coarse-grained forms of the non-bonding and bending interactions, with energy scales that serve as proxies for $\varepsilon_\mathrm{LJ}$ and $\varepsilon_\mathrm{bend}$, respectively~\cite{li2013challenges,salerno2016resolving}.
Our results here can then serve to guide the synthesis of materials by predicting whether the mechanical response will be in the nonmonotonic region, based on the  value of the control parameter $\varepsilon_\mathrm{bend}/\varepsilon_\mathrm{LJ}$.

\paragraph*{Outlook}
Nonmonotonic dependence of polymer glass mechanical properties results from two contrasting effects as polymer chain bending stiffness is increased: decreased density (and coordination) as monomer-monomer bending constraints are added; and increased mechanical rigidity of the chains.
Our results strongly support this being a general phenomenon, as it is robust all the way up to the glass transition temperature and persists for various sets of model parameters. Since bead-spring models form the basis of much contemporary theory for polymer glasses and their material properties, this finding has broad consequences across polymer physics.
Indeed, nonmonotonicity of dynamic quantities with respect to chain length and stiffness is emerging as a widespread feature of polymeric systems in various contexts~\cite{coronel2017non,bianco2016non}. 
It is, so far, difficult to isolate bending stiffness experimentally, since many other factors can influence the mechanical properties. Model systems such as colloidal and granular polymers (CGPs)~\cite{brown2012strain} might be good candidates for verifying our predicted nonmonotonicity, though, as they offer a very high level of control over coarse-grained properties.

The density of vibrational states from which we constructed the theoretical calculation of $G'$ using nonaffine lattice dynamics promises to offer additional insights into the structural and dynamic properties of polymer glasses in future works, both under shear induced yielding~\cite{rottler2001yield,rottler2003shear} and approaching $T^*_g$, and under imposed pressure~\cite{lin2016evidence}. Future work might extend the present finding to coarse-grained potentials that represent more specific materials~\cite{yi2013molecular,rosch2013exploring,salerno2016resolving}. Moreover, the present result represents the limit of long chains, while future work might explore the minimum chain length required to observe nonmonotonicity. This is further relevant to colloidal gels, where specific adhesive forces have been shown to lead to bending moments among small aggregates~\cite{pantina2005elasticity} that could influence the rheological properties~\cite{colombo2014stress} in an analogous way to that discussed here. Indeed, returning to the analogy with granular materials, it is not clear whether similar nonmonotonicity in $G'$ might be observed experimentally for increasing particle-particle friction. Recent theory~\cite{degiuli2017friction} suggests otherwise, as endogenous noise generated in such packings is responsible for rapidly opening and closing contacts meaning both the friction coefficient and $Z$ are rather poorly defined.
Understanding the role of rigidity in the mechanical properties of polymers will be useful in applications as diverse as packing genetic material in cells~\cite{rapaport2016packaging}, the structure of polyelectrolyte aggregates~\cite{C7SM01196B} and high-rate deformation of advanced materials~\cite{sirk2013high}.

\paragraph*{Acknowledgements}
CN acknowledges the Maudslay-Butler Research Fellowship at Pembroke College, Cambridge for financial support; VVP and AZ acknowledge financial support from the US Army Research Laboratory under grant nr. W911NF-16-2-0091.


\begin{thebibliography}{42}%
\makeatletter
\providecommand \@ifxundefined [1]{%
 \@ifx{#1\undefined}
}%
\providecommand \@ifnum [1]{%
 \ifnum #1\expandafter \@firstoftwo
 \else \expandafter \@secondoftwo
 \fi
}%
\providecommand \@ifx [1]{%
 \ifx #1\expandafter \@firstoftwo
 \else \expandafter \@secondoftwo
 \fi
}%
\providecommand \natexlab [1]{#1}%
\providecommand \enquote  [1]{``#1''}%
\providecommand \bibnamefont  [1]{#1}%
\providecommand \bibfnamefont [1]{#1}%
\providecommand \citenamefont [1]{#1}%
\providecommand \href@noop [0]{\@secondoftwo}%
\providecommand \href [0]{\begingroup \@sanitize@url \@href}%
\providecommand \@href[1]{\@@startlink{#1}\@@href}%
\providecommand \@@href[1]{\endgroup#1\@@endlink}%
\providecommand \@sanitize@url [0]{\catcode `\\12\catcode `\$12\catcode
  `\&12\catcode `\#12\catcode `\^12\catcode `\_12\catcode `\%12\relax}%
\providecommand \@@startlink[1]{}%
\providecommand \@@endlink[0]{}%
\providecommand \url  [0]{\begingroup\@sanitize@url \@url }%
\providecommand \@url [1]{\endgroup\@href {#1}{\urlprefix }}%
\providecommand \urlprefix  [0]{URL }%
\providecommand \Eprint [0]{\href }%
\providecommand \doibase [0]{http://dx.doi.org/}%
\providecommand \selectlanguage [0]{\@gobble}%
\providecommand \bibinfo  [0]{\@secondoftwo}%
\providecommand \bibfield  [0]{\@secondoftwo}%
\providecommand \translation [1]{[#1]}%
\providecommand \BibitemOpen [0]{}%
\providecommand \bibitemStop [0]{}%
\providecommand \bibitemNoStop [0]{.\EOS\space}%
\providecommand \EOS [0]{\spacefactor3000\relax}%
\providecommand \BibitemShut  [1]{\csname bibitem#1\endcsname}%
\let\auto@bib@innerbib\@empty
\bibitem [{\citenamefont {Doi}\ and\ \citenamefont
  {Edwards}(1988)}]{doi1988theory}%
  \BibitemOpen
  \bibfield  {author} {\bibinfo {author} {\bibfnamefont {M.}~\bibnamefont
  {Doi}}\ and\ \bibinfo {author} {\bibfnamefont {S.~F.}\ \bibnamefont
  {Edwards}},\ }\href@noop {} {\emph {\bibinfo {title} {The Theory of Polymer
  Dynamics}}},\ Vol.~\bibinfo {volume} {73}\ (\bibinfo  {publisher} {Oxford
  University Press, Oxford},\ \bibinfo {year} {1988})\BibitemShut {NoStop}%
\bibitem [{\citenamefont {Rubinstein}\ and\ \citenamefont
  {Colby}(2003)}]{Rubinstein2003}%
  \BibitemOpen
  \bibfield  {author} {\bibinfo {author} {\bibfnamefont {M.}~\bibnamefont
  {Rubinstein}}\ and\ \bibinfo {author} {\bibfnamefont {R.~H.}\ \bibnamefont
  {Colby}},\ }\href {\doibase 10.1017/CBO9780511975691} {\emph {\bibinfo
  {title} {Polymer Physics}}}\ (\bibinfo  {publisher} {Oxford University Press,
  Oxford},\ \bibinfo {year} {2003})\BibitemShut {NoStop}%
\bibitem [{\citenamefont {Hoy}(2011)}]{hoy2011understanding}%
  \BibitemOpen
  \bibfield  {author} {\bibinfo {author} {\bibfnamefont {R.~S.}\ \bibnamefont
  {Hoy}},\ }\href@noop {} {\bibfield  {journal} {\bibinfo  {journal} {Journal
  of Polymer Science Part B: Polymer Physics}\ }\textbf {\bibinfo {volume}
  {49}},\ \bibinfo {pages} {979} (\bibinfo {year} {2011})}\BibitemShut
  {NoStop}%
\bibitem [{\citenamefont {He}\ and\ \citenamefont
  {Thorpe}(1985)}]{he1985elastic}%
  \BibitemOpen
  \bibfield  {author} {\bibinfo {author} {\bibfnamefont {H.}~\bibnamefont
  {He}}\ and\ \bibinfo {author} {\bibfnamefont {M.~F.}\ \bibnamefont
  {Thorpe}},\ }\href@noop {} {\bibfield  {journal} {\bibinfo  {journal}
  {Physical Review Letters}\ }\textbf {\bibinfo {volume} {54}},\ \bibinfo
  {pages} {2107} (\bibinfo {year} {1985})}\BibitemShut {NoStop}%
\bibitem [{\citenamefont {Song}\ \emph {et~al.}(2008)\citenamefont {Song},
  \citenamefont {Wang},\ and\ \citenamefont {Makse}}]{song2008phase}%
  \BibitemOpen
  \bibfield  {author} {\bibinfo {author} {\bibfnamefont {C.}~\bibnamefont
  {Song}}, \bibinfo {author} {\bibfnamefont {P.}~\bibnamefont {Wang}}, \ and\
  \bibinfo {author} {\bibfnamefont {H.~A.}\ \bibnamefont {Makse}},\ }\href@noop
  {} {\bibfield  {journal} {\bibinfo  {journal} {Nature}\ }\textbf {\bibinfo
  {volume} {453}},\ \bibinfo {pages} {629} (\bibinfo {year}
  {2008})}\BibitemShut {NoStop}%
\bibitem [{\citenamefont {Wyart}\ and\ \citenamefont
  {Cates}(2014)}]{wyart2014discontinuous}%
  \BibitemOpen
  \bibfield  {author} {\bibinfo {author} {\bibfnamefont {M.}~\bibnamefont
  {Wyart}}\ and\ \bibinfo {author} {\bibfnamefont {M.}~\bibnamefont {Cates}},\
  }\href@noop {} {\bibfield  {journal} {\bibinfo  {journal} {Physical Review
  Letters}\ }\textbf {\bibinfo {volume} {112}},\ \bibinfo {pages} {098302}
  (\bibinfo {year} {2014})}\BibitemShut {NoStop}%
\bibitem [{\citenamefont {Rosch}\ \emph {et~al.}(2013)\citenamefont {Rosch},
  \citenamefont {Brennan}, \citenamefont {Izvekov},\ and\ \citenamefont
  {Andzelm}}]{rosch2013exploring}%
  \BibitemOpen
  \bibfield  {author} {\bibinfo {author} {\bibfnamefont {T.~W.}\ \bibnamefont
  {Rosch}}, \bibinfo {author} {\bibfnamefont {J.~K.}\ \bibnamefont {Brennan}},
  \bibinfo {author} {\bibfnamefont {S.}~\bibnamefont {Izvekov}}, \ and\
  \bibinfo {author} {\bibfnamefont {J.~W.}\ \bibnamefont {Andzelm}},\
  }\href@noop {} {\bibfield  {journal} {\bibinfo  {journal} {Physical Review
  E}\ }\textbf {\bibinfo {volume} {87}},\ \bibinfo {pages} {042606} (\bibinfo
  {year} {2013})}\BibitemShut {NoStop}%
\bibitem [{\citenamefont {Zou}\ \emph {et~al.}(2009)\citenamefont {Zou},
  \citenamefont {Cheng}, \citenamefont {Rivers}, \citenamefont {Jaeger},\ and\
  \citenamefont {Nagel}}]{zou2009packing}%
  \BibitemOpen
  \bibfield  {author} {\bibinfo {author} {\bibfnamefont {L.-N.}\ \bibnamefont
  {Zou}}, \bibinfo {author} {\bibfnamefont {X.}~\bibnamefont {Cheng}}, \bibinfo
  {author} {\bibfnamefont {M.~L.}\ \bibnamefont {Rivers}}, \bibinfo {author}
  {\bibfnamefont {H.~M.}\ \bibnamefont {Jaeger}}, \ and\ \bibinfo {author}
  {\bibfnamefont {S.~R.}\ \bibnamefont {Nagel}},\ }\href@noop {} {\bibfield
  {journal} {\bibinfo  {journal} {Science}\ }\textbf {\bibinfo {volume}
  {326}},\ \bibinfo {pages} {408} (\bibinfo {year} {2009})}\BibitemShut
  {NoStop}%
\bibitem [{\citenamefont {Karayiannis}\ \emph {et~al.}(2009)\citenamefont
  {Karayiannis}, \citenamefont {Foteinopoulou},\ and\ \citenamefont
  {Laso}}]{karayiannis2009contact}%
  \BibitemOpen
  \bibfield  {author} {\bibinfo {author} {\bibfnamefont {N.~C.}\ \bibnamefont
  {Karayiannis}}, \bibinfo {author} {\bibfnamefont {K.}~\bibnamefont
  {Foteinopoulou}}, \ and\ \bibinfo {author} {\bibfnamefont {M.}~\bibnamefont
  {Laso}},\ }\href@noop {} {\bibfield  {journal} {\bibinfo  {journal} {Physical
  Review E}\ }\textbf {\bibinfo {volume} {80}},\ \bibinfo {pages} {011307}
  (\bibinfo {year} {2009})}\BibitemShut {NoStop}%
\bibitem [{\citenamefont {Lopatina}\ \emph {et~al.}(2011)\citenamefont
  {Lopatina}, \citenamefont {Reichhardt},\ and\ \citenamefont
  {Reichhardt}}]{lopatina2011jamming}%
  \BibitemOpen
  \bibfield  {author} {\bibinfo {author} {\bibfnamefont {L.}~\bibnamefont
  {Lopatina}}, \bibinfo {author} {\bibfnamefont {C.~O.}\ \bibnamefont
  {Reichhardt}}, \ and\ \bibinfo {author} {\bibfnamefont {C.}~\bibnamefont
  {Reichhardt}},\ }\href@noop {} {\bibfield  {journal} {\bibinfo  {journal}
  {Physical Review E}\ }\textbf {\bibinfo {volume} {84}},\ \bibinfo {pages}
  {011303} (\bibinfo {year} {2011})}\BibitemShut {NoStop}%
\bibitem [{\citenamefont {Hoy}(2017)}]{hoy2017jamming}%
  \BibitemOpen
  \bibfield  {author} {\bibinfo {author} {\bibfnamefont {R.~S.}\ \bibnamefont
  {Hoy}},\ }\href@noop {} {\bibfield  {journal} {\bibinfo  {journal} {Physical
  Review Letters}\ }\textbf {\bibinfo {volume} {118}},\ \bibinfo {pages}
  {068002} (\bibinfo {year} {2017})}\BibitemShut {NoStop}%
\bibitem [{\citenamefont {Papakonstantopoulos}\ and\ \citenamefont
  {de~Pablo}(2016)}]{papakonstantopoulos2016controlling}%
  \BibitemOpen
  \bibfield  {author} {\bibinfo {author} {\bibfnamefont {G.~J.}\ \bibnamefont
  {Papakonstantopoulos}}\ and\ \bibinfo {author} {\bibfnamefont {J.~J.}\
  \bibnamefont {de~Pablo}},\ }\href@noop {} {\bibfield  {journal} {\bibinfo
  {journal} {arXiv preprint arXiv:1610.03806}\ } (\bibinfo {year}
  {2016})}\BibitemShut {NoStop}%
\bibitem [{\citenamefont {Stukowski}(2009)}]{stukowski2009visualization}%
  \BibitemOpen
  \bibfield  {author} {\bibinfo {author} {\bibfnamefont {A.}~\bibnamefont
  {Stukowski}},\ }\href@noop {} {\bibfield  {journal} {\bibinfo  {journal}
  {Modelling and Simulation in Materials Science and Engineering}\ }\textbf
  {\bibinfo {volume} {18}},\ \bibinfo {pages} {015012} (\bibinfo {year}
  {2009})}\BibitemShut {NoStop}%
\bibitem [{\citenamefont {Plimpton}(1995)}]{plimpton1995fast}%
  \BibitemOpen
  \bibfield  {author} {\bibinfo {author} {\bibfnamefont {S.}~\bibnamefont
  {Plimpton}},\ }\href@noop {} {\bibfield  {journal} {\bibinfo  {journal}
  {Journal of Computational Physics}\ }\textbf {\bibinfo {volume} {117}},\
  \bibinfo {pages} {1} (\bibinfo {year} {1995})}\BibitemShut {NoStop}%
\bibitem [{\citenamefont {Kremer}\ and\ \citenamefont
  {Grest}(1990)}]{kremer1990dynamics}%
  \BibitemOpen
  \bibfield  {author} {\bibinfo {author} {\bibfnamefont {K.}~\bibnamefont
  {Kremer}}\ and\ \bibinfo {author} {\bibfnamefont {G.~S.}\ \bibnamefont
  {Grest}},\ }\href@noop {} {\bibfield  {journal} {\bibinfo  {journal} {The
  Journal of Chemical Physics}\ }\textbf {\bibinfo {volume} {92}},\ \bibinfo
  {pages} {5057} (\bibinfo {year} {1990})}\BibitemShut {NoStop}%
\bibitem [{Note1()}]{Note1}%
  \BibitemOpen
  \bibinfo {note} {The bidispersity introduced by differing rest lengths of LJ
  and FENE is sufficient to suppress crystallisation throughout~\cite
  {zaccarelli2009crystallization,nguyen2015effect}}\BibitemShut {NoStop}%
\bibitem [{Note2()}]{Note2}%
  \BibitemOpen
  \bibinfo {note} {A comprehensive description of our simulation model is
  provided in the Supplementary Material}\BibitemShut {NoStop}%
\bibitem [{\citenamefont {Han}\ \emph {et~al.}(1994)\citenamefont {Han},
  \citenamefont {Gee},\ and\ \citenamefont {Boyd}}]{han1994glass}%
  \BibitemOpen
  \bibfield  {author} {\bibinfo {author} {\bibfnamefont {J.}~\bibnamefont
  {Han}}, \bibinfo {author} {\bibfnamefont {R.~H.}\ \bibnamefont {Gee}}, \ and\
  \bibinfo {author} {\bibfnamefont {R.~H.}\ \bibnamefont {Boyd}},\ }\href@noop
  {} {\bibfield  {journal} {\bibinfo  {journal} {Macromolecules}\ }\textbf
  {\bibinfo {volume} {27}},\ \bibinfo {pages} {7781} (\bibinfo {year}
  {1994})}\BibitemShut {NoStop}%
\bibitem [{\citenamefont {Strobl}(1997)}]{strobl1997physics}%
  \BibitemOpen
  \bibfield  {author} {\bibinfo {author} {\bibfnamefont {G.~R.}\ \bibnamefont
  {Strobl}},\ }\href@noop {} {\emph {\bibinfo {title} {The Physics of
  Polymers}}},\ Vol.~\bibinfo {volume} {2}\ (\bibinfo  {publisher} {Springer,
  United Kingdom},\ \bibinfo {year} {1997})\BibitemShut {NoStop}%
\bibitem [{\citenamefont {Zaccone}\ and\ \citenamefont
  {Terentjev}(2013)}]{zaccone2013disorder}%
  \BibitemOpen
  \bibfield  {author} {\bibinfo {author} {\bibfnamefont {A.}~\bibnamefont
  {Zaccone}}\ and\ \bibinfo {author} {\bibfnamefont {E.~M.}\ \bibnamefont
  {Terentjev}},\ }\href@noop {} {\bibfield  {journal} {\bibinfo  {journal}
  {Physical Review Letters}\ }\textbf {\bibinfo {volume} {110}},\ \bibinfo
  {pages} {178002} (\bibinfo {year} {2013})}\BibitemShut {NoStop}%
\bibitem [{\citenamefont {Lema{\^\i}tre}\ and\ \citenamefont
  {Maloney}(2006)}]{lemaitre2006sum}%
  \BibitemOpen
  \bibfield  {author} {\bibinfo {author} {\bibfnamefont {A.}~\bibnamefont
  {Lema{\^\i}tre}}\ and\ \bibinfo {author} {\bibfnamefont {C.}~\bibnamefont
  {Maloney}},\ }\href@noop {} {\bibfield  {journal} {\bibinfo  {journal}
  {Journal of Statistical Physics}\ }\textbf {\bibinfo {volume} {123}},\
  \bibinfo {pages} {415} (\bibinfo {year} {2006})}\BibitemShut {NoStop}%
\bibitem [{\citenamefont {Zaccone}\ and\ \citenamefont
  {Scossa-Romano}(2011)}]{zaccone2011approximate}%
  \BibitemOpen
  \bibfield  {author} {\bibinfo {author} {\bibfnamefont {A.}~\bibnamefont
  {Zaccone}}\ and\ \bibinfo {author} {\bibfnamefont {E.}~\bibnamefont
  {Scossa-Romano}},\ }\href@noop {} {\bibfield  {journal} {\bibinfo  {journal}
  {Physical Review B}\ }\textbf {\bibinfo {volume} {83}},\ \bibinfo {pages}
  {184205} (\bibinfo {year} {2011})}\BibitemShut {NoStop}%
\bibitem [{\citenamefont {Damart}\ \emph {et~al.}(2017)\citenamefont {Damart},
  \citenamefont {Tanguy},\ and\ \citenamefont {Rodney}}]{damart2017theory}%
  \BibitemOpen
  \bibfield  {author} {\bibinfo {author} {\bibfnamefont {T.}~\bibnamefont
  {Damart}}, \bibinfo {author} {\bibfnamefont {A.}~\bibnamefont {Tanguy}}, \
  and\ \bibinfo {author} {\bibfnamefont {D.}~\bibnamefont {Rodney}},\
  }\href@noop {} {\bibfield  {journal} {\bibinfo  {journal} {Physical Review
  B}\ }\textbf {\bibinfo {volume} {95}},\ \bibinfo {pages} {054203} (\bibinfo
  {year} {2017})}\BibitemShut {NoStop}%
\bibitem [{\citenamefont {Born}\ and\ \citenamefont
  {Huang}(1954)}]{born1954dynamical}%
  \BibitemOpen
  \bibfield  {author} {\bibinfo {author} {\bibfnamefont {M.}~\bibnamefont
  {Born}}\ and\ \bibinfo {author} {\bibfnamefont {K.}~\bibnamefont {Huang}},\
  }\href@noop {} {\emph {\bibinfo {title} {Dynamical theory of crystal
  lattices}}}\ (\bibinfo  {publisher} {Clarendon press, Oxford},\ \bibinfo
  {year} {1954})\BibitemShut {NoStop}%
\bibitem [{\citenamefont {Van~Workum}\ \emph {et~al.}(2006)\citenamefont
  {Van~Workum}, \citenamefont {Gao}, \citenamefont {Schall},\ and\
  \citenamefont {Harrison}}]{van2006expressions}%
  \BibitemOpen
  \bibfield  {author} {\bibinfo {author} {\bibfnamefont {K.}~\bibnamefont
  {Van~Workum}}, \bibinfo {author} {\bibfnamefont {G.}~\bibnamefont {Gao}},
  \bibinfo {author} {\bibfnamefont {J.~D.}\ \bibnamefont {Schall}}, \ and\
  \bibinfo {author} {\bibfnamefont {J.~A.}\ \bibnamefont {Harrison}},\
  }\href@noop {} {\bibfield  {journal} {\bibinfo  {journal} {Journal of
  Chemical Physics}\ }\textbf {\bibinfo {volume} {125}},\ \bibinfo {pages}
  {144506} (\bibinfo {year} {2006})}\BibitemShut {NoStop}%
\bibitem [{\citenamefont {Li}\ \emph {et~al.}(2013)\citenamefont {Li},
  \citenamefont {Abberton}, \citenamefont {Kr{\"o}ger},\ and\ \citenamefont
  {Liu}}]{li2013challenges}%
  \BibitemOpen
  \bibfield  {author} {\bibinfo {author} {\bibfnamefont {Y.}~\bibnamefont
  {Li}}, \bibinfo {author} {\bibfnamefont {B.~C.}\ \bibnamefont {Abberton}},
  \bibinfo {author} {\bibfnamefont {M.}~\bibnamefont {Kr{\"o}ger}}, \ and\
  \bibinfo {author} {\bibfnamefont {W.~K.}\ \bibnamefont {Liu}},\ }\href@noop
  {} {\bibfield  {journal} {\bibinfo  {journal} {Polymers}\ }\textbf {\bibinfo
  {volume} {5}},\ \bibinfo {pages} {751} (\bibinfo {year} {2013})}\BibitemShut
  {NoStop}%
\bibitem [{\citenamefont {Salerno}\ \emph {et~al.}(2016)\citenamefont
  {Salerno}, \citenamefont {Agrawal}, \citenamefont {Perahia},\ and\
  \citenamefont {Grest}}]{salerno2016resolving}%
  \BibitemOpen
  \bibfield  {author} {\bibinfo {author} {\bibfnamefont {K.~M.}\ \bibnamefont
  {Salerno}}, \bibinfo {author} {\bibfnamefont {A.}~\bibnamefont {Agrawal}},
  \bibinfo {author} {\bibfnamefont {D.}~\bibnamefont {Perahia}}, \ and\
  \bibinfo {author} {\bibfnamefont {G.~S.}\ \bibnamefont {Grest}},\ }\href@noop
  {} {\bibfield  {journal} {\bibinfo  {journal} {Physical Review Letters}\
  }\textbf {\bibinfo {volume} {116}},\ \bibinfo {pages} {058302} (\bibinfo
  {year} {2016})}\BibitemShut {NoStop}%
\bibitem [{\citenamefont {Coronel}\ \emph {et~al.}(2017)\citenamefont
  {Coronel}, \citenamefont {Orlandini},\ and\ \citenamefont
  {Micheletti}}]{coronel2017non}%
  \BibitemOpen
  \bibfield  {author} {\bibinfo {author} {\bibfnamefont {L.}~\bibnamefont
  {Coronel}}, \bibinfo {author} {\bibfnamefont {E.}~\bibnamefont {Orlandini}},
  \ and\ \bibinfo {author} {\bibfnamefont {C.}~\bibnamefont {Micheletti}},\
  }\href@noop {} {\bibfield  {journal} {\bibinfo  {journal} {Soft Matter}\
  }\textbf {\bibinfo {volume} {13}},\ \bibinfo {pages} {4260} (\bibinfo {year}
  {2017})}\BibitemShut {NoStop}%
\bibitem [{\citenamefont {Bianco}\ and\ \citenamefont
  {Malgaretti}(2016)}]{bianco2016non}%
  \BibitemOpen
  \bibfield  {author} {\bibinfo {author} {\bibfnamefont {V.}~\bibnamefont
  {Bianco}}\ and\ \bibinfo {author} {\bibfnamefont {P.}~\bibnamefont
  {Malgaretti}},\ }\href@noop {} {\bibfield  {journal} {\bibinfo  {journal}
  {The Journal of Chemical Physics}\ }\textbf {\bibinfo {volume} {145}},\
  \bibinfo {pages} {114904} (\bibinfo {year} {2016})}\BibitemShut {NoStop}%
\bibitem [{\citenamefont {Brown}\ \emph {et~al.}(2012)\citenamefont {Brown},
  \citenamefont {Nasto}, \citenamefont {Athanassiadis},\ and\ \citenamefont
  {Jaeger}}]{brown2012strain}%
  \BibitemOpen
  \bibfield  {author} {\bibinfo {author} {\bibfnamefont {E.}~\bibnamefont
  {Brown}}, \bibinfo {author} {\bibfnamefont {A.}~\bibnamefont {Nasto}},
  \bibinfo {author} {\bibfnamefont {A.~G.}\ \bibnamefont {Athanassiadis}}, \
  and\ \bibinfo {author} {\bibfnamefont {H.~M.}\ \bibnamefont {Jaeger}},\
  }\href {\doibase 10.1103/PhysRevLett.108.108302} {\bibfield  {journal}
  {\bibinfo  {journal} {Phys. Rev. Lett.}\ }\textbf {\bibinfo {volume} {108}},\
  \bibinfo {pages} {108302} (\bibinfo {year} {2012})}\BibitemShut {NoStop}%
\bibitem [{\citenamefont {Rottler}\ and\ \citenamefont
  {Robbins}(2001)}]{rottler2001yield}%
  \BibitemOpen
  \bibfield  {author} {\bibinfo {author} {\bibfnamefont {J.}~\bibnamefont
  {Rottler}}\ and\ \bibinfo {author} {\bibfnamefont {M.~O.}\ \bibnamefont
  {Robbins}},\ }\href@noop {} {\bibfield  {journal} {\bibinfo  {journal}
  {Physical Review E}\ }\textbf {\bibinfo {volume} {64}},\ \bibinfo {pages}
  {051801} (\bibinfo {year} {2001})}\BibitemShut {NoStop}%
\bibitem [{\citenamefont {Rottler}\ and\ \citenamefont
  {Robbins}(2003)}]{rottler2003shear}%
  \BibitemOpen
  \bibfield  {author} {\bibinfo {author} {\bibfnamefont {J.}~\bibnamefont
  {Rottler}}\ and\ \bibinfo {author} {\bibfnamefont {M.~O.}\ \bibnamefont
  {Robbins}},\ }\href@noop {} {\bibfield  {journal} {\bibinfo  {journal}
  {Physical Review E}\ }\textbf {\bibinfo {volume} {68}},\ \bibinfo {pages}
  {011507} (\bibinfo {year} {2003})}\BibitemShut {NoStop}%
\bibitem [{\citenamefont {Lin}\ \emph {et~al.}(2016)\citenamefont {Lin},
  \citenamefont {Jorjadze}, \citenamefont {Pontani}, \citenamefont {Wyart},\
  and\ \citenamefont {Brujic}}]{lin2016evidence}%
  \BibitemOpen
  \bibfield  {author} {\bibinfo {author} {\bibfnamefont {J.}~\bibnamefont
  {Lin}}, \bibinfo {author} {\bibfnamefont {I.}~\bibnamefont {Jorjadze}},
  \bibinfo {author} {\bibfnamefont {L.-L.}\ \bibnamefont {Pontani}}, \bibinfo
  {author} {\bibfnamefont {M.}~\bibnamefont {Wyart}}, \ and\ \bibinfo {author}
  {\bibfnamefont {J.}~\bibnamefont {Brujic}},\ }\href@noop {} {\bibfield
  {journal} {\bibinfo  {journal} {Physical Review Letters}\ }\textbf {\bibinfo
  {volume} {117}},\ \bibinfo {pages} {208001} (\bibinfo {year}
  {2016})}\BibitemShut {NoStop}%
\bibitem [{\citenamefont {Yi}\ \emph {et~al.}(2013)\citenamefont {Yi},
  \citenamefont {Locker},\ and\ \citenamefont {Rutledge}}]{yi2013molecular}%
  \BibitemOpen
  \bibfield  {author} {\bibinfo {author} {\bibfnamefont {P.}~\bibnamefont
  {Yi}}, \bibinfo {author} {\bibfnamefont {C.~R.}\ \bibnamefont {Locker}}, \
  and\ \bibinfo {author} {\bibfnamefont {G.~C.}\ \bibnamefont {Rutledge}},\
  }\href@noop {} {\bibfield  {journal} {\bibinfo  {journal} {Macromolecules}\
  }\textbf {\bibinfo {volume} {46}},\ \bibinfo {pages} {4723} (\bibinfo {year}
  {2013})}\BibitemShut {NoStop}%
\bibitem [{\citenamefont {Pantina}\ and\ \citenamefont
  {Furst}(2005)}]{pantina2005elasticity}%
  \BibitemOpen
  \bibfield  {author} {\bibinfo {author} {\bibfnamefont {J.~P.}\ \bibnamefont
  {Pantina}}\ and\ \bibinfo {author} {\bibfnamefont {E.~M.}\ \bibnamefont
  {Furst}},\ }\href@noop {} {\bibfield  {journal} {\bibinfo  {journal}
  {Physical Review Letters}\ }\textbf {\bibinfo {volume} {94}},\ \bibinfo
  {pages} {138301} (\bibinfo {year} {2005})}\BibitemShut {NoStop}%
\bibitem [{\citenamefont {Colombo}\ and\ \citenamefont
  {Del~Gado}(2014)}]{colombo2014stress}%
  \BibitemOpen
  \bibfield  {author} {\bibinfo {author} {\bibfnamefont {J.}~\bibnamefont
  {Colombo}}\ and\ \bibinfo {author} {\bibfnamefont {E.}~\bibnamefont
  {Del~Gado}},\ }\href@noop {} {\bibfield  {journal} {\bibinfo  {journal}
  {Journal of Rheology}\ }\textbf {\bibinfo {volume} {58}},\ \bibinfo {pages}
  {1089} (\bibinfo {year} {2014})}\BibitemShut {NoStop}%
\bibitem [{\citenamefont {DeGiuli}\ and\ \citenamefont
  {Wyart}(2017)}]{degiuli2017friction}%
  \BibitemOpen
  \bibfield  {author} {\bibinfo {author} {\bibfnamefont {E.}~\bibnamefont
  {DeGiuli}}\ and\ \bibinfo {author} {\bibfnamefont {M.}~\bibnamefont
  {Wyart}},\ }\href@noop {} {\bibfield  {journal} {\bibinfo  {journal} {arXiv
  preprint arXiv:1704.00740}\ } (\bibinfo {year} {2017})}\BibitemShut {NoStop}%
\bibitem [{\citenamefont {Rapaport}(2016)}]{rapaport2016packaging}%
  \BibitemOpen
  \bibfield  {author} {\bibinfo {author} {\bibfnamefont {D.}~\bibnamefont
  {Rapaport}},\ }\href@noop {} {\bibfield  {journal} {\bibinfo  {journal}
  {Physical Review E}\ }\textbf {\bibinfo {volume} {94}},\ \bibinfo {pages}
  {030401} (\bibinfo {year} {2016})}\BibitemShut {NoStop}%
\bibitem [{\citenamefont {Mima}\ \emph {et~al.}(2017)\citenamefont {Mima},
  \citenamefont {Kinjo}, \citenamefont {Yamakawa},\ and\ \citenamefont
  {Asahi}}]{C7SM01196B}%
  \BibitemOpen
  \bibfield  {author} {\bibinfo {author} {\bibfnamefont {T.}~\bibnamefont
  {Mima}}, \bibinfo {author} {\bibfnamefont {T.}~\bibnamefont {Kinjo}},
  \bibinfo {author} {\bibfnamefont {S.}~\bibnamefont {Yamakawa}}, \ and\
  \bibinfo {author} {\bibfnamefont {R.}~\bibnamefont {Asahi}},\ }\href
  {\doibase 10.1039/C7SM01196B} {\bibfield  {journal} {\bibinfo  {journal}
  {Soft Matter}\,\ } (\bibinfo {year} {2017})}\BibitemShut {NoStop}%
\bibitem [{\citenamefont {Sirk}\ \emph {et~al.}(2013)\citenamefont {Sirk},
  \citenamefont {Khare}, \citenamefont {Karim}, \citenamefont {Lenhart},
  \citenamefont {Andzelm}, \citenamefont {McKenna},\ and\ \citenamefont
  {Khare}}]{sirk2013high}%
  \BibitemOpen
  \bibfield  {author} {\bibinfo {author} {\bibfnamefont {T.~W.}\ \bibnamefont
  {Sirk}}, \bibinfo {author} {\bibfnamefont {K.~S.}\ \bibnamefont {Khare}},
  \bibinfo {author} {\bibfnamefont {M.}~\bibnamefont {Karim}}, \bibinfo
  {author} {\bibfnamefont {J.~L.}\ \bibnamefont {Lenhart}}, \bibinfo {author}
  {\bibfnamefont {J.~W.}\ \bibnamefont {Andzelm}}, \bibinfo {author}
  {\bibfnamefont {G.~B.}\ \bibnamefont {McKenna}}, \ and\ \bibinfo {author}
  {\bibfnamefont {R.}~\bibnamefont {Khare}},\ }\href@noop {} {\bibfield
  {journal} {\bibinfo  {journal} {Polymer}\ }\textbf {\bibinfo {volume} {54}},\
  \bibinfo {pages} {7048} (\bibinfo {year} {2013})}\BibitemShut {NoStop}%
\bibitem [{\citenamefont {Zaccarelli}\ \emph {et~al.}(2009)\citenamefont
  {Zaccarelli}, \citenamefont {Valeriani}, \citenamefont {Sanz}, \citenamefont
  {Poon}, \citenamefont {Cates},\ and\ \citenamefont
  {Pusey}}]{zaccarelli2009crystallization}%
  \BibitemOpen
  \bibfield  {author} {\bibinfo {author} {\bibfnamefont {E.}~\bibnamefont
  {Zaccarelli}}, \bibinfo {author} {\bibfnamefont {C.}~\bibnamefont
  {Valeriani}}, \bibinfo {author} {\bibfnamefont {E.}~\bibnamefont {Sanz}},
  \bibinfo {author} {\bibfnamefont {W.}~\bibnamefont {Poon}}, \bibinfo {author}
  {\bibfnamefont {M.}~\bibnamefont {Cates}}, \ and\ \bibinfo {author}
  {\bibfnamefont {P.}~\bibnamefont {Pusey}},\ }\href@noop {} {\bibfield
  {journal} {\bibinfo  {journal} {Physical Review Letters}\ }\textbf {\bibinfo
  {volume} {103}},\ \bibinfo {pages} {135704} (\bibinfo {year}
  {2009})}\BibitemShut {NoStop}%
\bibitem [{\citenamefont {Nguyen}\ \emph {et~al.}(2015)\citenamefont {Nguyen},
  \citenamefont {Smith}, \citenamefont {Hoy},\ and\ \citenamefont
  {Karayiannis}}]{nguyen2015effect}%
  \BibitemOpen
  \bibfield  {author} {\bibinfo {author} {\bibfnamefont {H.~T.}\ \bibnamefont
  {Nguyen}}, \bibinfo {author} {\bibfnamefont {T.~B.}\ \bibnamefont {Smith}},
  \bibinfo {author} {\bibfnamefont {R.~S.}\ \bibnamefont {Hoy}}, \ and\
  \bibinfo {author} {\bibfnamefont {N.~C.}\ \bibnamefont {Karayiannis}},\
  }\href@noop {} {\bibfield  {journal} {\bibinfo  {journal} {Journal of
  Chemical Physics}\ }\textbf {\bibinfo {volume} {143}},\ \bibinfo {pages}
  {144901} (\bibinfo {year} {2015})}\BibitemShut {NoStop}%
\end{thebibliography}

%

\end{document}